# Tunneling spectroscopy of c-axis epitaxial cuprate junctions

Panpan Zhou,[1] Liyang Chen,[2] Ilya Sochnikov,[3] Tsz Chun Wu,[1] Matthew S. Foster,[1] Anthony T. Bollinger,[4] Xi He,[4,5] Ivan Božović,[4,5*] and Douglas Natelson[1,6,7*]

[1]Department of Physics and Astronomy, Rice University, Houston, TX 77005, USA
[2]Applied Physics Graduate Program, Smalley-Curl Institute, Rice University, Houston, TX 77005, USA
[3]Department of Physics, University of Connecticut, Storrs, CT 06269, USA
[4]Brookhaven National Laboratory, Upton, New York 11973-5000, USA
[5]Department of Chemistry, Yale University, New Haven CT 06520, USA
[6]Department of Electrical and Computer Engineering, Rice University, Houston, TX 77005 USA
[7]Department of Materials Science and NanoEngineering, Rice University, Houston, TX 77005 USA

(Dated: January 8, 2020)

## Abstract

Atomically precise epitaxial structures are unique systems for tunneling spectroscopy that minimize extrinsic effects of disorder. We present a systematic tunneling spectroscopy study, over a broad doping, temperature, and bias range, in epitaxial $c$-axis La$_{2-x}$Sr$_x$CuO$_4$/La$_2$CuO$_4$/La$_{2-x}$Sr$_x$CuO$_4$ heterostructures. The behavior of these superconductor/insulator/superconductor (SIS) devices is unusual. Down to 20 mK there is complete suppression of $c$-axis Josephson critical current with a barrier of only 2 nm of La$_2$CuO$_4$, and the zero-bias conductance remains at 20-30% of the normal-state conductance, implying a substantial population of in-gap states. Tunneling spectra show greatly suppressed coherence peaks. As the temperature is raised, the superconducting gap fills in rather than closing at $T_c$. For all doping levels, the spectra show an inelastic tunneling feature at $\sim$ 80 meV, suppressed as $T$ exceeds $T_c$. These nominally simple epitaxial cuprate junctions deviate markedly from expectations based on the standard Bardeen-Cooper-Schrieffer (BCS) theory.

Tunneling spectroscopy has proven to be an important tool in studying superconducting materials. In conventional superconductors, the tunneling spectroscopy of normal metal-insulator-superconductor (NIS) junctions confirms the Bardeen-Cooper-Schrieffer (BCS) form of the single-particle density of states (DOS) [1] and provides a direct measurement of the superconducting gap as a function of temperature [2]. Spectral features have also revealed the presence of inelastic tunneling processes [3]. In superconductor-insulator superconductor (SIS) junctions made from the conventional low-temperature superconductors, features in the inelastic tunneling spectrum provided a quantitative demonstration that the pairing originates from electron-phonon interaction [4].

Tunneling spectroscopy in cuprate high-temperature superconductors in recent years has largely employed scanning tunneling spectroscopy [5]. Thanks to the spatial resolution of scanning tunneling microscopy (STM), this provides a means of assessing the local density of states with atomic resolution. Such measurements have demonstrated spatial heterogeneity in the magnitude of the superconducting gap [6–8] and provided data on the relation between pseudogap and superconductivity [9]. In-depth study of spatial correlations in local tunneling spectra has revealed signatures of other ordered states[10], and momentum-space information through quasiparticle interference spectroscopy[11].

With state-of-the-art atomic layer-by-layer molecular beam epitaxy (ALL-MBE) [12], it is possible to fabricate $c$-axis copper-oxide trilayer heterostructures with atomically flat interfaces and minimal disorder (limited by the nanoscale distribution of dopant atoms). SIS junctions have been demonstrated using $La_{2-x}Sr_xCuO_4$ (LSCO) as the superconducting top and bottom electrodes, with the intervening tunnel barrier consisting of the undoped, antiferromagnetic, Mott-insulator parent compound $La_2CuO_4$ (LCO) [13]. The high-quality and uniformity of these trilayer heterostructures has been proven by extremely narrow superconducting transitions observed in mutual inductance measurements[14], and by demonstrating that merely one-unit-cell thick LCO barriers are sufficient to inhibit superconducting current between the LSCO electrodes, implying the absence of pinholes[13]. Shot noise in the tunneling current in such junctions indicates the presence of pair charge carriers both above $T_c$ and at energies large compared to the superconducting gap scale [15].

Here we present a systematic study of the tunneling characteristics of LSCO/LCO/LSCO epitaxial tunnel junctions, spanning a broad range of temperature, bias, and doping. Consistent with prior work involving *c*-axis tunneling through LCO barriers [13], these devices show no signs of a coherent Josephson supercurrent down to the lowest temperatures (20 mK) and currents (~ 3 picoamperes) measured. Different from conventional Josephson junctions, the LSCO/LCO/LSCO junctions show a vanishingly small critical current-normal state resistance product, $I_cR_N$ (at least 6 orders of magnitude smaller than $2\Delta/e$, where $\Delta$ is the nominal superconducting gap inferred from bias-dependent suppression of the tunneling conductance), despite remarkable structural order. Below $T_c$ of the superconducting electrode LSCO films, we find strongly suppressed coherence peaks as well as large residual conductance; even for $T\rightarrow 0$, the latter remains typically at about 30% the normal-state ($T > T_c$) differential conductance. The tunneling characteristics are not consistent with the expectations for planar *c*-axis *d*-wave BCS SIS tunneling. The gap fills in rather than closing as $T$ is increased above $T_c$, as in the phenomenological "Dynes superconductor" model [16, 17]. Inelastic features are also present in the tunneling conductance, but these are suppressed as $T$ is increased above $T_c$.

The LSCO/LCO/LSCO trilayer films were grown using the ALL-MBE system. The film was deposited on LaSrAlO$_4$ (LSAO) substrates and the growth process was monitored and controlled in real-time by reflection high-energy electron diffraction (RHEED). A source of pure ozone is used to ensure sufficient oxidation under high-vacuum conditions. The substrate temperature was kept at 650°C and the ozone partial pressure at $2\times10^{-5}$ Torr. Atomic-resolution scanning transmission electron microscopy (STEM) and energy-dispersive x-ray spectroscopy (EDEX) demonstrate atomically sharp interfaces and remarkable crystalline perfection. Both top and bottom LSCO layers show an extremely narrow superconducting transition, confirming the uniformity and high structural quality of the films [14].

The tunneling devices were fabricated from the LSCO/LCO/LSCO films using photolithography techniques. After photolithography to define mesa locations, the film was milled down to the substrate with argon ions into 20 μm square mesas. A second lithography step defined circular tunnel junctions, with a second controlled ion milling to etch the surrounding material through the top LSCO layer and the middle LCO layer, exposing (but not etching through) the bottom LSCO layer. To isolate the top and bottom Au contacts, a thick layer of $Al_2O_3$ was evaporated to photolithographically defined areas. Finally, Au is evaporated to make contact with top/bottom LSCO layers.[16]

The device electrical properties were measured with standard lock-in amplifier techniques. The measurements were performed from room temperature down to 2 K in a variable temperature cryostat, and down to 20 mK in a separate measurement setup within a dilution refrigerator. The $R(T)$ curves show that the device conductance is dominated by the insulating LCO layer. Devices with doping level $x = 0.10, 0.12, 0.14$ and $0.15$ (close to optimum doping) in LSCO electrodes showed the superconducting transition temperatures at 28 K, 34 K, 37 K and 38 K, respectively, consistent with our other LSCO films and the literature. For each doping, we performed differential conductance measurements on multiple devices over a broad range of voltage bias and temperatures.

The superconducting critical current $I_c$ is unmeasurably small for all the samples down to the lowest temperatures. For traditional SIS Josephson tunneling junctions, the $I_cR_N$ product is expected to be comparable to the gap voltage scale [18]. The complete suppression of $I_cR_N$ seen in these junctions remains remarkable. Prior measurements have revealed that the coherence length for this type of tunnel junction is very short along the $c$-axis [12], so that even a 1.5 unit cell (2 nm) barrier is empirically sufficient to prevent supercurrent between the upper and lower LSCO layers. (In contrast, in analogous structures with underdoped LSCO barriers, long-range proximity-induced supercurrent has been observed through very underdoped $La_2CuO_{4+\delta}$ barriers as thick as 46 nm [19, 20].) The reproducibility of $R_N$ and the lack of a measurable $I_c$ through 2 nm of LCO indicate the high-quality of these junctions, the lack of parasitic conduction around the junction perimeter, the absence of pinholes, and that the undoped insulator is extremely effective at suppressing $c$-axis supercurrent. Since undoped LCO is an antiferromagnetic Mott insulator in bulk, it is worth considering whether the magnetic degrees of freedom[21] in the 1.5 unit cell thick LCO barrier might play an important role in this suppression.

For all temperatures, the differential conductance has a V-shaped background in the normal state, consistent with the pseudogap, that extends from above the transition temperature $T_c$ of each film, to the superconducting temperature regime (Fig. 1). There is an overall asymmetry to d$I$/d$V$ vs. $V_{dc}$, with the conductance being higher at the polarity such that electrons are driven from the bottom LSCO layer to the top. This asymmetry is consistent with the gradient in epitaxial strain away from the substrate, and the polar nature of the material [22]. At temperatures below $T_c$, the conductance is very nonlinear and exhibits a suppression at zero bias, as expected for a SIS junction. The zero-bias conductance suppression becomes progressively sharper as the doping is varied from near-optimal $x = 0.15$ to the more underdoped $x = 0.10$. For all devices, as $T \to 0$ the differential conductance at zero bias $dI/dV(V_{dc} = 0)$ saturates at a finite value rather than extrapolating to zero. This saturation implies the presence of a large population of in-gap states in the LSCO/LCO/LSCO even as $T \to 0$.

Figure 1 shows the differential conductance tunneling spectra of representative devices for the four doping levels in LSCO. The suppression of the low-bias conductance below $T_c$ is apparent, as is the residual zero-bias conductance. We consider the functional form of these tunneling spectra below. The naive expectation for a structurally clean, large-area tunnel junction is the conservation of crystal momentum in the $a-b$ plane. However, a calculation based on a BCS order parameter and transverse $k$ conservation, for planar tunneling of perfectly 2D quasiparticles, is in strong disagreement with the experimental data.

Figure 2 shows normalized tunneling spectra, $(dI/dV(V,T))/(dI/dV(V,T = 50$ K$))$, a rough attempt to focus on the superconducting gap aspects of SIS tunneling while minimizing the role of the higher energy pseudogap and inherent device asymmetry. The data do not conform to the standard BCS expectations. We have attempted a phenomenological approach by fitting to the $d$-wave version of the Dynes formula [23] for the tunneling density of states in presence of strong in-plane lifetime and scattering effects that affect the self-energy:

$$N(\omega) = N_0 \, \text{Re} \left\langle \left[ \frac{\omega + i\Gamma(\omega,T)}{\sqrt{(\omega + i\Gamma(\omega,T))^2 - \Delta^2 \cos^2(2\theta)}} \right] \right\rangle_\theta \quad (1)$$

where $N_0$ is an overall normalization, ω is energy, $\Gamma$ is the effective lifetime broadening, $\Delta$ is the magnitude of the *d*-wave gap, and $2\theta$ describes the angular dependence of the gap within the *a-b* plane. Strictly speaking, the addition of any $\Gamma(\omega,T)$ is a deviation from standard BCS theory. We have used an ansatz $\Gamma(\omega,T) = \alpha(T)\omega + \beta(T)$ that has been employed in interpreting STM tunneling spectra in cuprates[24]. The "standard" Dynes approach, with a frequency-independent contribution $\beta(T)$ often introduced as a pair-breaking rate [17,25], does not fit the data well, being unable to balance the suppression of coherence peaks and the residual zero-bias conductance. A contribution $\alpha(T)\omega$ could arise from the scattering of nodal Dirac quasiparticles (as expected in the *d*-wave cuprates) from quenched disorder [26]. The expected differential conductance ignoring kinetic constraints on transverse *k* is then

$$\frac{dI}{dV} = A\frac{d}{dV}\int N(\omega+eV)N(\omega)[f(\omega)-f(\omega+eV)]d\omega \quad (2)$$

where *V* is the dc bias voltage, *f* is the Fermi-Dirac distribution function, and the prefactor *A* accounts for normalization. The relaxation of the constraint of transverse *k* conservation can result from multiple reasons — $k_z$ dispersion, spatial inhomogeneity of the electronic structure in the LSCO electrodes (as seen in STM spectra in other cuprates[24]), for example.

Using the equation (2), we try to fit the differential conductance with the $\alpha(T)$, $\beta(T)$ and $\Delta(T)$ as the fitting parameters. The model works relatively well for the $x = 0.15$ doped sample, as shown in Figure 3. The temperature dependence of the fit parameters $\alpha(T)$, $\beta(T)$ and $\Delta(T)$ is as expected for Dynes superconductors. As *T* increases toward $T_c$, the gap seems to 'fill in' due to an increasing $\Gamma$, rather than 'closing' due to decreasing $\Delta$. This is consistent with observations made in photoemission experiments of other cuprates [27, 28]. $\alpha(T)$ has a weak temperature dependence and $\beta(T)$ has a quadratic temperature dependence, as the black dash line in Figure 3b indicates. The $\Gamma$ broadening both suppresses coherence peaks and, through $\beta$, leads to residual $T = 0$ conductance. Fitting with this choice of $\Gamma(\omega,T)$ is not satisfactory for the more underdoped samples, however. The primary difficulty remains in achieving a proper balance between the suppression of the coherence peaks and residual zero-bias conductivity as $T \to 0$. Note, too, that fitting is sensitive to the $T$-normalization procedure, which means that any temperature evolution of the pseudogap could distort the normalized data and affect the fitting results. The difficulties in fitting the conductance spectra argue for the need for further theoretical examination of SIS tunneling in such systems, including the roles of disorder and the LCO barrier.

The second derivative of the $I(V)$ characteristics, $d^2I/dV^2$, reveals inelastic tunneling features. Fig. 4 shows inelastic tunneling analysis as a function of temperature for the various LSCO doping levels. Numerical differentiation of the differential conductance, $dI/dV$, is quantitatively consistent with the directly measured lock-in second harmonic signal, $d^2I/dV^2$. To isolate inelastic features, panels (e)-(h) plot the derivative of the symmetrized conductance, $(1/2)(dI/dV(+V)+dI/dV(-V))$.[29] For all the devices, there are broad inelastic features at energies at around 0.08 eV that become markedly weaker as $T$ is increased and become undetectable above $T_c$, appearing to decrease in magnitude rather than broadening or shifting to lower biases. With increasing of doping levels from 0.10 to 0.15, the inelastic features become less prominent. At higher biases exceeding 300 meV (not shown), strong shot noise and device instabilities make it more difficult to resolve inelastic features, if any, at higher energies.

In SIS junctions, tunneling of quasiparticles via a coupling to a bosonic mode of energy $\varepsilon$ manifests in a $d^2I/dV^2$ feature at bias $eV = \varepsilon + 2\Delta$. [30] Depending on the relative importance of elastic and inelastic processes involving the mode, one can expect a dip (for dominant elastic corrections to the self-energy) or a peak (for inelastic tunneling contributing an additional channel for conduction) in $d^2I/dV^2$.[31] Inferring the gap from the width of the $dI/dV$ suppression, the relevant bosonic mode energy in this case is around 65 meV. This is close to the experimentally observed out-of-plane oxygen vibrations (∼55 meV) known to couple strongly to the carriers [32, 33], and B1g and half-breathing modes in LCO found by neutron scattering in this energy range [34]. The voltage width of the $d^2I/dV^2$ feature is comparable to the width of the distribution of inelastic tunneling feature positions observed by STM in BSCCO[35].

In summary, we have performed a tunneling spectroscopy study of LSCO/LCO/LSCO tunnel junctions from the underdoped to near optimal doping, revealing several marked deviations from BCS expectations. Despite the high structural perfection inherent in epitaxially grown structures, the SIS tunneling spectra are better fit by a form that omits constraints on the transverse momentum. A phenomenological Dynes model can account for strong suppression of coherence peaks and large residual zero-bias conduction at $T{\to}0$, indicating a large contribution of in-gap states even far below $T_c$. This is coincident with maximal violation of the conventional Ambegaokar-Baratoff relationship between $I_c$ and $R_N$. The complete suppression of $I_c$ occurs with only a 2 nm LCO barrier, despite the facts that shot-noise measurements [14] indicate the presence of a pair contribution to the tunneling transport, and that underdoped LSCO barriers show robust long-ranged proximity-induced superconductivity [19, 20]. Inelastic tunneling spectra below $T_c$ reveal features in an energy range near to that of the known phonon modes; these features are suppressed above $T_c$. Further studies of such epitaxial junctions, particularly in the presence of large magnetic fields and different combinations of doping levels and barrier structures, should shed further light on the tunneling process, the role of the LCO barrier, and the nature of relevant bosonic modes.

**Acknowledgements.** The authors acknowledge C. Hooley and A. Nevidomskyy for useful discussions at early stages of this work, as well as Joerg Schmalian. The research at Brookhaven National Laboratory, including heterostructure synthesis and characterization and device fabrication, was supported by the US Department of Energy, Basic Energy Sciences, Materials Sciences and Engineering Division. X.H. was supported by the Gordon and Betty Moore Foundation's EPiQS Initiative through grant GBMF4410. The work at the University of Connecticut was supported by the US state of Connecticut. The experimental research at Rice University was supported by the US Department of Energy, Basic Energy Sciences, Experimental Condensed Matter Physics award DE-FG02-06ER46337. M. Foster and T.C. Wu acknowledge support from Welch Foundation Grant No. C-1809.

* bozovic@bnl.gov

* natelson@rice.edu

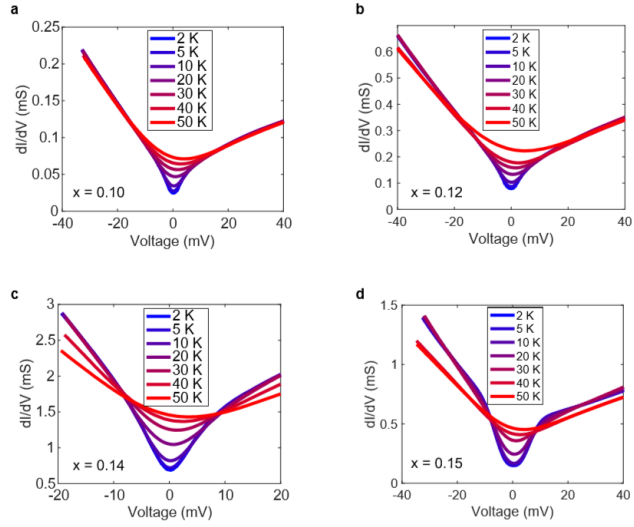

FIG. 1. Differential conductance $dI/dV$ as a function of $V_{dc}$ for the doping levels $x$ = 0.10, 0.12, 0.14, and 0.15, in panels (a-d), respectively. The bias asymmetry correlates with the structure of the junctions, while the broader V-shape is a manifestation of the pseudogap.

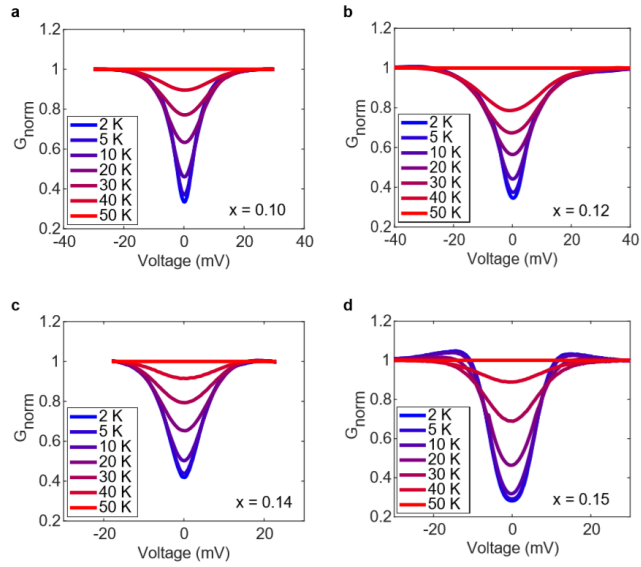

FIG. 2. Normalized differential conductance $G_{norm} \equiv (dI/dV)/(dI/dV(T = 50\ K))$, for the doping levels $x$ = 0.10, 0.12, 0.14, and 0.15 in panels (a-d), respectively.

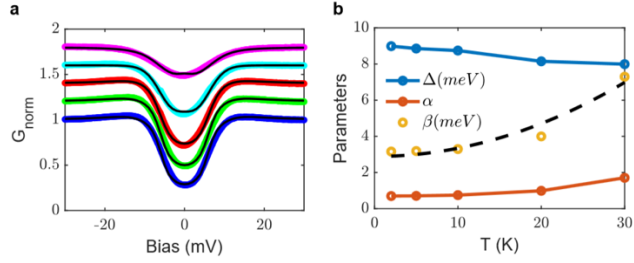

FIG. 3. Fitting to the normalized differential conductance and the corresponding fitting parameters at various temperatures below $T_c$. (a) Normalized differential conductance for doping $x = 0.15$ at 2 K (blue), 5 K (green), 10 K (red), 20 K (cyan) and 30 K (magenta). The black lines are the fittings to the conductance at each temperature. Data are shifted by 0.2 vertically between each temperature. (b) The fitting parameters $\Delta$, $\alpha$ and $\beta$ as a function of temperature.

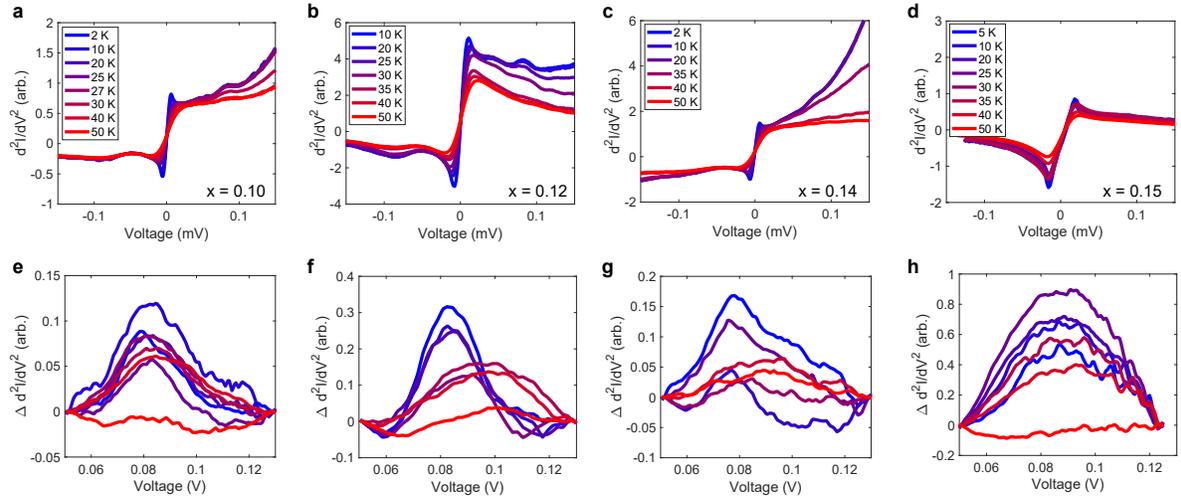

FIG. 4. (a-d) Inelastic spectra, d$^2I$/d$V^2$ as a function of bias for $x$ = 0.10, 0.12, 0.14, and 0.15, respectively. The fine solid line is a fit of the lowest temperature data to Eqs. (1) and (2). (e-h) Close-up views of the positive polarity part of the antisymmetrized inelastic tunneling spectra, with a smooth polynomial background (obtained at 50 K) subtracted.
1616